\documentclass[aps,twocolumn,prl,showpacs,amsmath,amssymb,10pt]{revtex4-1}
\usepackage{graphicx}
\usepackage{scrextend}
\usepackage{manfnt,pifont}
\usepackage{hyperref}
\usepackage{url}

\usepackage{url}

\usepackage{CJK}

\begin{document}

\begin{CJK*}{GB}{}
\CJKfamily{gbsn}

\title{Super Gluon Five-Point Amplitudes in AdS Space}

\author{Luis F. Alday}
\affiliation{Mathematical Institute, University of Oxford, Andrew Wiles Building, Radcliffe Observatory Quarter, Woodstock Road, Oxford, OX2 6GG, U.K.}
\author{Vasco Gon\c{c}alves}
\affiliation{Centro de F\'isica do Porto e Departamento de F\'isica e Astronomia, Faculdade de Ci\^encias da Universidade do Porto, Rua do Campo Alegre 687, 4169-007 Porto, Portugal.}
\author{Xinan Zhou (ÖÜÏ¡éª)}
\affiliation{Kavli Institute for Theoretical Sciences, University of Chinese Academy of Sciences, Beijing 100190, China.}

\begin{abstract}
\noindent We present the tree-level five-point amplitude of the lowest KK mode of SYM on AdS$_5$$\times$S$^3$, dual to the correlator of the flavor current multiplet in the dual 4d ${\cal N}=2$ SCFT. Its color and kinematical structure is particularly simple and resembles that of the flat-space gluon amplitude.

\end{abstract}

\maketitle
\end{CJK*}

\section{Introduction}
\noindent
Via the AdS/CFT correspondence, correlation functions in CFTs are identified with on-shell scattering amplitudes in AdS space. On the one hand, from these holographic correlators we can extract useful data of the strongly coupled CFTs. On the other, we can use them to explore generalizations of various properties of flat-space amplitudes in curved backgrounds, which might hopefully shed light on the yet-elusive underlying principles. 

Recent explorations have led to encouraging results in the search for AdS avatars of flat-space properties. For example, there has been much evidence for AdS realizations of the color-kinematic duality \cite{Farrow:2018yni,Lipstein:2019mpu,Armstrong:2020woi,Albayrak:2020fyp,Alday:2021odx,Diwakar:2021juk} and a double copy relation has been identified in AdS$_5$ \cite{Zhou:2021gnu}. Both properties play a pivotal role in the modern flat-space amplitude program and have numerous applications \footnote{See \cite{Bern:2019prr} for a review.}. However, it should be noted that most of progress in AdS has been limited to correlators with $n\leq 4$ points. For a systematic understanding of these properties it is  imperative to study higher-point functions. 

In this paper, we present the first five-point amplitude of AdS super gluons. In particular, we consider the lowest Kaluza-Klein (KK) modes of SYM on AdS$_5$$\times$S$^3$, which arises as a decoupling sector in the holographic dual of several important 4d $\mathcal{N}=2$ SCFTs \cite{Fayyazuddin:1998fb,Aharony:1998xz,Karch:2002sh}. Compared to supergravity, the kinematics of super gluons is much simpler, making them the ideal arena for higher-point explorations. On the other hand, an eventual full-fledge  AdS double copy relation would also make gluon amplitudes more fundamental. Unfortunately, computing holographic correlators by traditional methods is in general very difficult. This is partially due to the complicated effective Lagrangian, and also because of the many curved-space diagrams needed to be summed up. Instead, we will apply the modern techniques of \cite{Rastelli:2016nze,Rastelli:2017udc,Zhou:2017zaw,Alday:2020lbp,Alday:2020dtb}, and in particular \cite{Goncalves:2019znr}, to bootstrap the correlator. We show that the five-point function can be fixed by using only symmetries and consistency conditions, without inputing precise details of the effective Lagrangian. Moreover, we find a remarkably simple expression for the super gluon Mellin amplitude 
\begin{equation}
\nonumber \mathcal{M}_5=\sum_{i=1}^{15}\frac{g_i n_i}{D_i}\;,
\end{equation}
which has the same form as the gluon five-point amplitude in flat space.  Here $g_i$ are color structures given by cubic graphs and $D_i$ are the associated scalar propagators. The kinematic numerators $n_i$ turn out to satisfy the same relations as the color factors $g_i$, therefore giving rise to a color-kinematic duality. While the color-ordered super gluon amplitudes satisfy Kleiss-Kuijf relations \cite{Kleiss:1988ne} and the photon-decoupling identity as their flat-space counterpart, we show that interestingly they do not obey BCJ relations \cite{Bern:2008qj}. Furthermore, we find that the corrections terms are particularly simple and hint at a pattern that may generalize to higher points.

\section{Kinematics}
\noindent
AdS super gluons are dual to scalar operators of the form $\mathcal{O}_k^{I;\alpha_1\ldots \alpha_k}$, labelled by an integer $k=2,3,\ldots$. They are the superconformal primaries of 
$\frac{1}{2}$-BPS multiplets of the 4d $\mathcal{N}=2$ superconformal algebra, with protected conformal dimension $\Delta=k$. These operators transform in the adjoint representation of a color group $G_F$ \footnote{Via the AdS/CFT correspondence, $G_F$ is a global symmetry from the CFT perspective.} and in the spin-$\frac{k}{2}$ representation of the $SU(2)_R$ R-symmetry group. In top-down holographic models \cite{Fayyazuddin:1998fb,Aharony:1998xz,Karch:2002sh}, they arise as the $S^3$ KK modes of an $\mathcal{N}=1$ vector multiplet on an AdS$_5$$\times$S$^3$ subspace. In the large central charge limit, their couplings with the gravitational degrees of freedom in the 10d ambient space are parametrically suppressed. Therefore, at leading order it is consistent to consider the AdS$_5$$\times$S$^3$ theory by itself. In this paper, we will focus on operators with the $k=2$ KK level. 

To conveniently keep track of the $SU(2)_R$ indices, we contract them with two-component polarization spinors $\mathcal{O}_2^I(x;v)\equiv\mathcal{O}_2^{I;\alpha_1,\alpha_2}(x)v_{\alpha_1}v_{\alpha_2}$. The five-point function 
\begin{equation}
G_5(x_i;v_i)=\langle \mathcal{O}_2^{I_1}(x_1;v_1)\ldots \mathcal{O}_2^{I_5}(x_5;v_5)\rangle
\end{equation}
becomes a function of both the spacetime and the internal coordinates. Here, we suppressed the color indices on the LHS to avoid cluttering the notation. Clearly, $G_5$ is a polynomial of $(v_i,v_j)\equiv \epsilon^{\alpha\beta}v_{i,\alpha}v_{j,\beta}$ and every $v_i$ should appear in each monomial precisely twice. Covariance under conformal symmetry and R-symmetry allows us to write $G_5$ in terms of cross ratios. Moreover, fermionic generators in the superconformal group impose additional constraints. In particular, the chiral algebra construction \cite{bllprv13} constrains the form of correlators when all operators are inserted on a 2d plane. Let us parameterize the plane by complex coordinates $(z,\bar{z})$. For the special coordinate-dependent polarizations $v_i^\alpha=(1,\bar{z}_i)$, the resulting correlator is independent of the anti-holomorphic coordinates \cite{bllprv13}
\begin{equation}\label{chiralalgebracond}
\partial_{\bar{z}_j}G_5\big(z_i,\bar{z}_i;v_i^\alpha=(1,\bar{z}_i)\big)=0\;.
\end{equation}

\begin{figure}[h]
\centering
\includegraphics[width=0.45\textwidth]{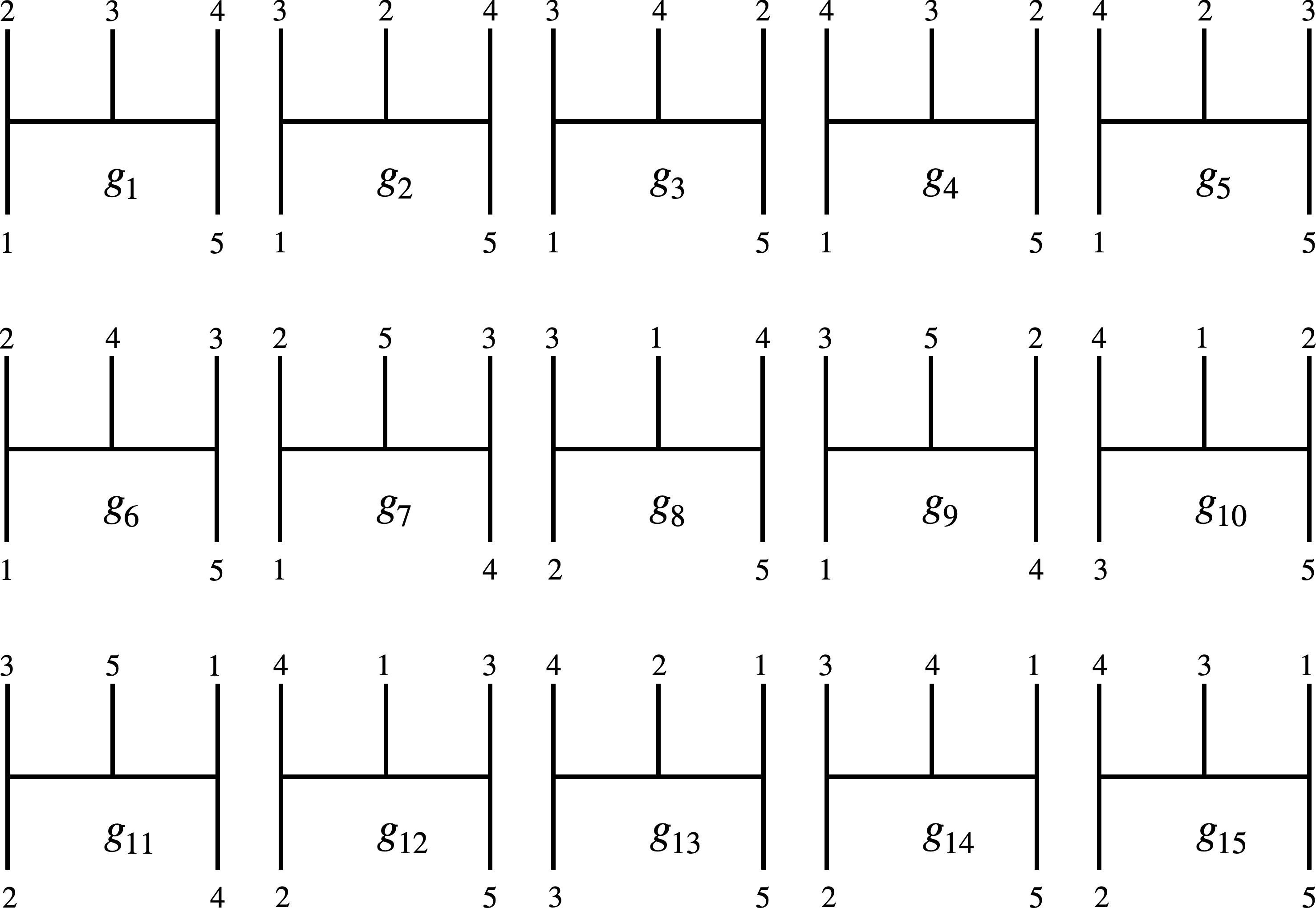}
\caption{All possible cubic tree diagrams.}
    \label{fig:cubicd}
\end{figure}

The five-point function has multiple color structures. In this paper we will focus on correlators which correspond to tree-level scattering in AdS. Moreover, all the particles are in the adjoint representation. Therefore, as in flat space, the color factors are linear combinations of the cubic tree diagrams $\{g_1,\ldots,g_{15}\}$ enumerated in fig. \ref{fig:cubicd} . Each diagram represents a contraction of the color group structure constants $f^{IJK}$, {\it e.g.},
\begin{equation}
g_1=f^{I_1I_2J}f^{JI_3K}f^{KI_4I_5}\;.
\end{equation}
These cubic diagrams are not all independent but are related by the Jacobi identity. For instance,
\begin{equation}\label{grelationsex}
g_1-g_2+g_8=0\;,\quad g_1-g_6+g_7=0\;,
\end{equation}
with the complete set of relations given in (\ref{grelations}). These relations reduce the number of independent $g_i$ to six. 

Finally, a convenient way to represent holographic correlators is the Mellin representation formalism \cite{Mack:2009mi,Penedones:2010ue} where the analytic structure is manifest. An $n$-point function can be written as
\begin{equation}
G_n(x_i)=\int [ds_{ij}] \bigg(\prod_{i<j} (x_{ij}^2)^{-s_{ij}}\Gamma[s_{ij}] \bigg)\,\mathcal{M}_n(s_{ij})
\end{equation}
where $\mathcal{M}(s_{ij})$ is the {\it Mellin amplitude}. The Mellin-Mandelstam variables $s_{ij}$ satisfy relations identical to those of the flat-space Mandelstam variables $s_{ij}^\flat$
\begin{equation}\label{sijrelations}
s_{ij}=s_{ji}\;,\quad s_{ii}=-\Delta_i\;, \quad \sum_{j=1}^n s_{ij}=0\;,
\end{equation}
except that the squared masses are replaced by the conformal dimensions. One can also recover the flat-space amplitude by taking the high energy limit \cite{Penedones:2010ue}
\begin{equation}\label{flatlimit}
\mathcal{T}^\flat_n(s^\flat_{ij}) \propto \lim_{\beta\to\infty} \beta^a \mathcal{M}_n(\beta s^\flat_{ij})\;,
\end{equation}
where $a$ is some appropriate power. But we note that the limit can be zero because the polarizations in $\mathcal{T}^\flat_n(s^\flat_{ij})$ are restricted to special configurations \cite{Alday:2020dtb}. 

\section{Bootstrap}
 \noindent
In principle, $G_5$ can be computed by standard diagrammatic techniques. However, this would require explicit details of the off-shell effective Lagrangian whose complexity obscures the simplicity of the final ``on-shell'' correlator. A more efficient strategy is to bootstrap the answer using supersymmetry, as pioneered in \cite{Rastelli:2016nze,Rastelli:2017udc}. Here we will use the method of \cite{Goncalves:2019znr} which is specifically tailored for five-point functions.

\begin{figure}[h]
\centering
\includegraphics[width=0.45\textwidth]{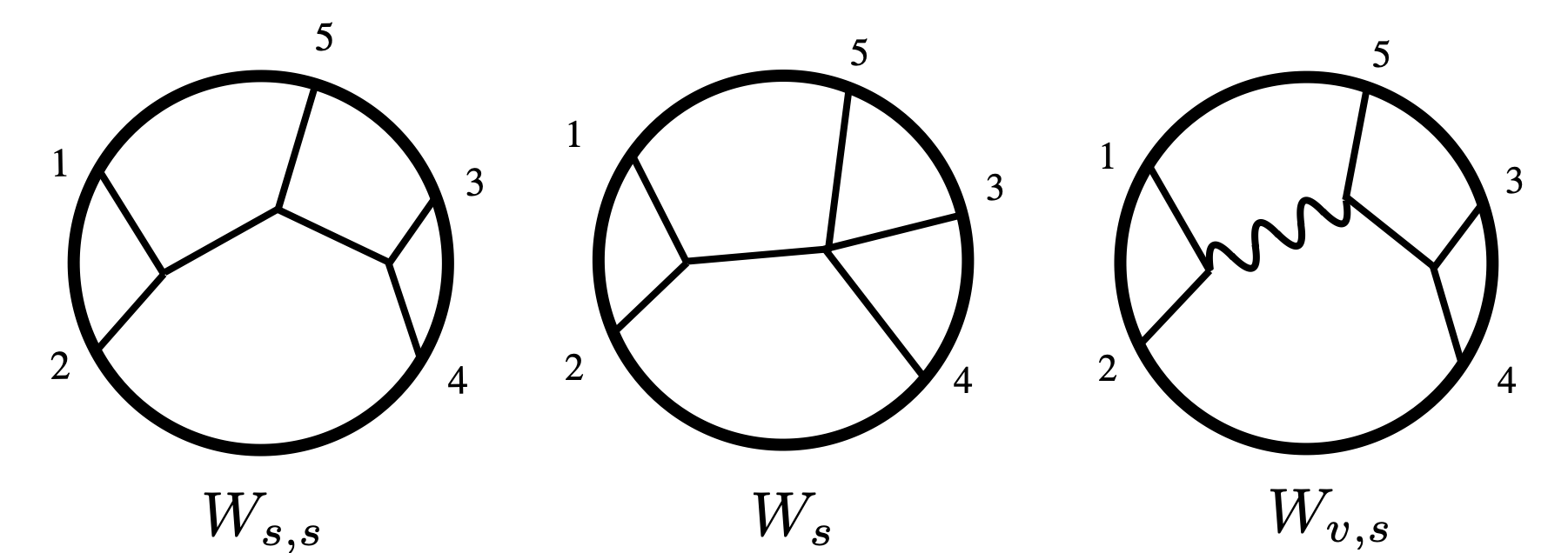}
\caption{Tree-level Witten diagrams. Straight lines represent scalar super gluons ($s$) and wavy lines are spin-1 gluons ($v$).}
    \label{fig:WDs}
\end{figure}

We start with an ansatz which is a linear combination of all possible Witten diagrams. The only possible exchanged fields are the scalar super gluon field itself and a spin-1 gluon field. The latter vector field is a superconformal descendant and therefore also carries the adjoint representation of the color group. It has conformal dimension $\Delta=3$ and is neutral under $SU(2)_R$. The ansatz is a linear combination of the exchange diagrams listed in fig. \ref{fig:WDs}
\begin{equation}\label{Gansatz}
\begin{split}
G_5^{\rm ansatz}={}&Y_{s,s}(g,v)W_{s,s}+Y_s(g,v)W_{s}\\
{}&+Y_{v,s}(g,v)W_{v,s}+({\rm permutations})\;.
\end{split}
\end{equation}
Here $Y_X(g,v)$ are color and R-symmetry structures compatible with the diagram, parameterized by unknown coefficients. For example, there is a unique color and R-symmetry structure associated with the first diagram
\begin{equation}
Y_{s,s}(g,v)=\lambda_{s,s} g_7(V_{13245}+V_{13254}+V_{13524}-V_{14235})
\end{equation} 
where $\lambda_{s,s}$ is an unknown parameter and $V_{abcde}$ are pentagon-shaped contractions of the polarization spinors
\begin{equation}
V_{abcde}=(v_a\cdot v_b)(v_b\cdot v_c)(v_c\cdot v_d)(v_d\cdot v_e)(v_e\cdot v_a)\;.
\end{equation}
The Witten diagrams $W_{s,s}$, $W_s$, $W_{v,s}$ were evaluated in \cite{Goncalves:2019znr} as finite linear combinations of $D$-functions $D_{\Delta_1,\Delta_2,\Delta_3,\Delta_4,\Delta_5}$. For example,
\begin{equation}\label{Wss}
W_{s,s}=\frac{1}{16x_{12}^2x_{34}^2}D_{1,1,1,1,2}\;.
\end{equation}
More details about R-symmetry structures and Witten diagrams can be found in the Supplemental Material. In arriving at the ansatz, we have also used symmetry and compatibility with the flat-space limit. R-symmetry prohibits double-exchange diagrams where both internal lines are spin-1 gluons. On the other hand, the polarizations in the flat-space limit $\mathcal{T}^\flat_5(s^\flat_{ij})$ are orthogonal to all momenta \cite{Alday:2020dtb}. This forces the flat-space limit to vanish \footnote{In flat space there are five polarization vectors. There must be at least one vector contracted with and a momentum, and gives zero due to transversality.}, forbiding vector single-exchange diagrams and five-point contact diagrams in the ansatz \footnote{In Mellin space, these diagrams will be leading in the high energy limit compared to the diagrams in the ansatz.}. In total, there are eight unknown coefficients in the ansatz (\ref{Gansatz}). Crossing symmetry reduces this number down to three. 

Finally, as was noted in \cite{Goncalves:2019znr} all $D$-functions in the ansatz are derivatives of $D_{1,1,1,1,2}$ (and its permutations) with respect to $x_{ij}^2$. On the other hand, the basic $D$-function $D_{1,1,1,1,2}$ was studied in the amplitude literature and can be expressed in terms of one-loop box diagrams \cite{Bern:1992em,Bern:1993kr}. Using these properties we can explicitly evaluate the $x_{ij}^2$-dependence of the ansatz, and implementing the chiral algebra condition (\ref{chiralalgebracond}) becomes straightforward. This condition is highly nontrivial and translates to an over-determined set of linear equations for the unknown parameters. We find that these equations fix the ansatz completely up to an overall constant.

\section{Five-point function}
\noindent
We now present the five-point function. The Mellin amplitude is remarkably simple and can be written in a form resembling the flat-space gluon amplitude
\begin{equation}\label{Mellin5pt}
\mathcal{M}_5=\sum_{i=1}^{15}\frac{g_i n_i}{D_i}\;.
\end{equation}
Here, $D_i$ are Mellin scalar propagators associated with the i-th diagram. More precisely, if the external labels of $g_i$ are clockwisely $(abcde)$ ({\it e.g.}, $g_1$ reads (12345)), then 
\begin{equation}
D_i=(s_{ab}-1)(s_{de}-1)\;.
\end{equation}
The kinematic factors $n_i$ are given by \footnote{Here we have chosen the coupling to be such that kinematic factors have this normalization.}
\begin{equation}
n_i=s_{ad} V_{acdeb}-s_{ae}V_{acedb}+s_{bd}V_{aedcb}-s_{be}V_{adecb}\;.
\end{equation}
Remarkably, the kinematic factors satisfy the same algebraic relations as the color factors $g_i$, {\it e.g.}, 
\begin{equation}
n_1-n_2+n_8=0\;,\quad n_1-n_6+n_7=0\;.
\end{equation}
This gives rise to an AdS version of the {\it color-kinematic duality} \cite{Bern:2008qj} at the level of five-point functions, and generalizes the previous observation at four points \cite{Alday:2021odx}. As a consistency check, (\ref{Mellin5pt}) also factorizes correctly into lower-point amplitudes at its poles \cite{Goncalves:2014rfa}. 

The Mellin expression (\ref{Mellin5pt}) also leads to a very compact representation of the five-point function in position space
\begin{equation}\label{G5posi}
G_5=-\frac{32}{\pi^2}\sum_{i=1}^{15}g_i \mathbb{D}_i W_i\;.
\end{equation}
Here, $W_i$ is the scalar double-exchange Witten diagram associated with $g_i$ ($W_{s,s}$ in (\ref{Wss}) up to permutations), and $\mathbb{D}_i$ are differential operators defined by
\begin{equation}
\mathbb{D}_i=\theta_{ad} V_{acdeb}-\theta_{ae}V_{acedb}+\theta_{bd}V_{aedcb}-\theta_{be}V_{adecb}\;,
\end{equation}
with $\theta_{ab}\equiv x_{ab}^2\frac{\partial}{\partial x_{ab}^2}$. We  checked that (\ref{G5posi}) is equivalent to the position space expression directly obtained from (\ref{Gansatz}), upon using $D$-function identities. Setting $v_i^\alpha=(1,\bar{z}_i)$ in the co-plane configuration, we find a holomorphic function as required by the chiral algebra construction
\begin{equation}
G_5\big(z_i,\bar{z}_i;v_i^\alpha=(1,\bar{z}_i)\big)=\frac{4}{5}\sum_{i=1}^{15}Z_ig_i
\end{equation}
where 
\begin{equation}
Z_i=[abcde]+[baced]-[bacde]-[abced]\;,
\end{equation}
with $[abcde]=\left(z_{ab}z_{bc}z_{cd}z_{de}z_{ea}\right)^{-1}$ and $z_{ij}=z_i-z_j$.

\section{Amplitude relations}
\noindent
The resemblance between flat-space and AdS amplitudes is clear from (\ref{Mellin5pt}). To further explore the analogy, we will study the partial amplitudes
\begin{equation}
\nonumber \mathcal{M}_5=\sum_{\sigma\in S_4} A_5(1,\sigma(2),\ldots,\sigma(5)){\rm Tr}(T^{I_1}T^{I_{\sigma(2)}}\ldots T^{I_{\sigma(5)}})\,.
\end{equation}
Since all particles are in the adjoint color representation, the Kleiss-Kuijf relations and the photon-decoupling identity are automatically satisfied. On the other hand, we will show that the color-ordered amplitudes do {\it not} obey the flat-space BCJ relations, even though the Mellin amplitude satisfies color-kinematic duality. Interestingly, the correction terms exhibit a very simple structure.  

Concretely, let us consider the following partial amplitudes in the Del Duca-Dixon-Maltoni (DDM) basis $\{g_1,\ldots,g_6\}$ \cite{DelDuca:1999rs}  
\begin{eqnarray}\label{DDMamplitudes}
\nonumber A_5(1,2,3,4,5)&=&\frac{n_1}{D_1}-\frac{n_7}{D_7}-\frac{n_8}{D_8}+\frac{n_{13}}{D_{13}}-\frac{n_{14}}{D_{14}}\,,\\
\nonumber A_5(1,3,2,4,5)&=&\frac{n_2}{D_2}+\frac{n_8}{D_8}-\frac{n_9}{D_9}+\frac{n_{14}}{D_{14}}+\frac{n_{15}}{D_{15}}\,,\\
\nonumber A_5(1,3,4,2,5)&=&\frac{n_3}{D_3}+\frac{n_9}{D_9}-\frac{n_{10}}{D_{10}}-\frac{n_{13}}{D_{13}}-\frac{n_{15}}{D_{15}}\,,\\
\nonumber A_5(1,4,3,2,5)&=&\frac{n_4}{D_4}+\frac{n_{10}}{D_{10}}-\frac{n_{11}}{D_{11}}+\frac{n_{13}}{D_{13}}-\frac{n_{14}}{D_{14}}\,,\\
\nonumber A_5(1,4,2,3,5)&=&\frac{n_5}{D_5}+\frac{n_{11}}{D_{11}}+\frac{n_{12}}{D_{12}}+\frac{n_{14}}{D_{14}}+\frac{n_{15}}{D_{15}}\,,\\
\nonumber A_5(1,2,4,3,5)&=&\frac{n_6}{D_6}+\frac{n_7}{D_7}-\frac{n_{12}}{D_{12}}-\frac{n_{13}}{D_{13}}-\frac{n_{15}}{D_{15}}\,.
\end{eqnarray}
After using the relations among the kinematic factors, we can express the RHS in terms of the six independent kinematic factors $n_{1}, \cdots n_6$ . In other words, the six amplitudes are related to $n_{1}, \cdots n_6$ by the multiplication of a matrix which is made of scalar propagators $D_i$. One could try to invert the matrix and solve for $n_{1}, \cdots n_6$  in terms of the partial amplitudes. In flat space, this is impossible because the matrix contains vectors with zero eigenvalues. Multiplying such a vector from the left gives rise to the BCJ relations \cite{Bern:2008qj} which are permutations of 
\begin{equation}\label{flatBCJ}
\begin{split}
{}&s^\flat_{12}A^\flat_5(1,2,3,4,5)+(s^\flat_{12}+s^\flat_{23})A^\flat_5(1,3,2,4,5)\\
{}&+(s^\flat_{12}+s^\flat_{23}+s^\flat_{24})A^\flat_5(1,3,4,2,5)=0\;,
\end{split}
\end{equation}
and reduce the number of independent amplitudes from six to two. Crucially, all flat-space particles are massless, {\it i.e.}, $s^\flat_{ii}=0$ and all poles are at zero. In AdS space, however, the matrix becomes {\it invertible} thanks to the different relations (\ref{sijrelations}) and the massive poles in $D_i$. Therefore, there are no BCJ relations and all six color-ordered amplitudes are independent. Let us comment here that the absence of BCJ relations should not raise concerns. In flat space, the non-invertibility of the matrix could be argued by noting that the kinematic factors are gauge-dependent while the color-ordered amplitudes are gauge-independent. Being able to express the former in terms of the latter would violate gauge invariance. This argument rests on the scattering particles being quanta of a spin-1 gauge field. Here, instead, we consider the scattering amplitudes of a scalar field. Gauge invariance is trivial and does not change the kinematic factors. Our statement of no BCJ relations also does not contradict the ``differential BCJ relations'' found in \cite{Diwakar:2021juk}. For their construction, it is important that the external particles are {\it massless} while in our case they are {\it massive}.

Let us now examine the analogue of (\ref{flatBCJ}) in AdS space. Elementary manipulations lead to
\begin{equation}\label{AdS5ptre}
\begin{split}
\nonumber {}&s_{12}A_5(1,2,3,4,5)+(s_{12}+s_{23})A_5(1,3,2,4,5)\\
{}&+(s_{12}+s_{23}+s_{24})A_5(1,3,4,2,5)=\frac{n_1}{D_1}+\frac{n_2}{D_2}+\frac{n_3}{D_3}\\
{}&-\frac{n_7}{D_7}+\frac{n_8}{D_8}+\frac{n_9}{D_9}-\frac{n_{10}}{D_{10}}-\frac{n_{13}}{D_{13}}+\frac{n_{14}}{D_{14}}-\frac{n_{15}}{D_{15}}\;.
\end{split}
\end{equation}
Note that the RHS is remarkably simple and consists of the same building blocks appearing in the color-ordered amplitudes (\ref{DDMamplitudes}) with $\pm 1$ coefficients. It is instructive to compare it with the four-point case computed in \cite{Alday:2021odx}
\begin{equation}
\begin{split}
\nonumber s_{12} A_4(1,2,3,4)+(s_{12}+s_{23})A_4(1,3,2,4)=\frac{n_s}{s-2}{}&\\
+\frac{n_t}{t-2}-\frac{n_u}{u-2}{}&\;,
\end{split}
\end{equation}
where we find the same interesting pattern. The color-ordered amplitudes are given by
\begin{equation}
\begin{split}
\nonumber A_4(1,2,3,4)={}&\frac{n_s}{s-2}-\frac{n_t}{t-2}\;,\\
A_4(1,3,2,4)={}&\frac{n_t}{t-2}-\frac{n_u}{u-2}\;,
\end{split}
\end{equation}
with $s+t+u=8$ and $n_s+n_t+n_u=0$\;.

\section{Conclusions and outlook}
\noindent
In this paper we computed the five-point super gluon amplitude of the lowest KK level by imposing symmetry constraints and consistency conditions. We found extremely simple closed-form expressions both in Mellin space and in position space. We expect that the techniques can be used to compute amplitudes of higher KK modes. It would also be interesting to see if the same strategy allow us to fix higher-point functions. On the other hand, we noted that there is a remarkable resemblance between the Mellin amplitude and the flat-space gluon amplitude. Therefore, it might be possible to borrow techniques from the flat-space amplitude program and develop more constructive methods ({\it e.g.}, along the lines of Britto-Cachazo-Feng-Witten recursion relations \cite{Britto:2005fq}). Let us also note that although (\ref{Mellin5pt}) displays color-kinematic duality, naively replacing the color factors with kinematic factors does not reproduce the five-point supergravity amplitude obtained in \cite{Goncalves:2019znr}. In fact, such a prescription already fails at four points. However, the recent work \cite{Zhou:2021gnu} pointed out that such an AdS double copy relation does exist for four-point functions at the level of superconformal {\it reduced} amplitudes (corresponding to stripping off supercharge delta functions in flat space). It would be interesting to extend the definition of reduced amplitudes to higher points and to check if the double copy structure persists. 

\vspace{0.3cm}
We thank Bo Feng, Song He, Fei Teng and Gang Yang for discussions. The work of L.F.A. is supported by funding from the European Research Council (ERC) under the European Union's Horizon 2020 Research and Innovation Programme (grant agreement No 787185) and by the STFC grant ST/T000864/1. V.G. is supported by Simons Foundation grants \#488637 (Simons collaboration on the non-perturbative bootstrap). Centro
de F\'isica do Porto is partially funded by Funda\c{c}\~ao para a Ci\^encia e Tecnologia (FCT)
under the grant UID04650-FCUP. The work of X.Z. is supported by starting funds from University of Chinese Academy of Sciences (UCAS) and from the Kavli Institute for Theoretical Sciences (KITS).

\appendix

\section{{\large \sc{Supplemental Material}}}

\subsection{Color factor relations}
The Jacobi identity at each internal line of the tree diagrams implies linear relations among the color structures $g_i$. For instance,
\begin{equation}
\nonumber  0 = (f^{12I}f^{I3J} +f^{13I}f^{IJ2}+f^{1JI} f^{I23}) f^{J45} =  g_1-g_2+g_8\,.
\end{equation}
The total list of relations is
\begin{eqnarray}\label{grelations}
\nonumber &&g_1-g_2+g_8=0,~g_1-g_6+g_7=0,~g_2 - g_3 + g_9=0,\\
\nonumber &&g_3 - g_4 + g_{10}=0,~
g_4 - g_5 + g_{11}=0,~g_5 - g_6 - g_{12} = 0,\\
\nonumber &&g_7 - g_{10} + g_{13}=0,~
g_8 + g_{11} - g_{14}=0,\\ 
 &&g_9 - g_{12} + g_{15}=0,~ g_{13} + g_{14} - g_{15}=0\;,
\end{eqnarray}
which leaves only six independent color factors.

\subsection{Details of the ansatz}
\subsubsection{Witten diagrams}
The $D$-functions are contact Witten diagrams in AdS defined by the integral
\begin{equation}
\nonumber D_{\Delta_1,\Delta_2,\Delta_3,\Delta_4,\Delta_5}=\int\frac{dz_0d^4\vec{z}}{z_0^5}\prod_{i=1}^5\left(\frac{z_0}{z_0^2+(\vec{z}-\vec{x}_1)^2}\right)^{\Delta_i}\;.
\end{equation}
The exchange Witten diagrams can be expressed as linear combination of $D$-functions \cite{Goncalves:2019znr}
\begin{eqnarray}
\nonumber W_{s,s}=&&\frac{1}{16x_{12}^2x_{34}^2}D_{1,1,1,1,2}\;,\\
\nonumber W_s=&&\frac{D_{1,1,2,2,2}}{4x_{12}^2}\;,\\
\nonumber W_{v,s}=&&-\frac{1}{8x_{12}^2x_{34}^2}\big(x_{24}^2D_{1,2,1,2,2}-2x_{25}^2D_{1,2,1,1,3}\\
\nonumber &&+x_{23}^2D_{1,2,2,1,2}
+2x_{15}^2D_{2,1,1,1,3}-x_{14}^2D_{2,1,1,2,2}\\
\nonumber &&-x_{13}^2D_{2,1,2,1,2}\big)\;.
\end{eqnarray}
The $D$-functions satisfy differential recursion relations
\begin{equation}
\nonumber D_{\Delta_1,\ldots,\Delta_i+1,\ldots,\Delta_j+1,\ldots,\Delta_5}=\frac{4-\sum_{a=1}^5\Delta_a}{2\Delta_i\Delta_j}\frac{\partial}{\partial x_{ij}^2}D_{\Delta_1,\ldots,\Delta_5}\;.
\end{equation}
Using these relations, we can reduce all $D$-function to the basic one $D_{1,1,1,1,2}$ (and its permutations), which is known as the pentagon integral in the amplitude literature \cite{Bern:1992em,Bern:1993kr}. This function can be expressed in terms of the well known one-loop box diagram which evaluates to dilogarithms of the cross ratios \cite{Usyukina:1992jd}. We refer the reader to Appendix D of \cite{Goncalves:2019znr} for a detailed summary of these results. 

\begin{figure}[h]
\centering
\includegraphics[width=0.2\textwidth]{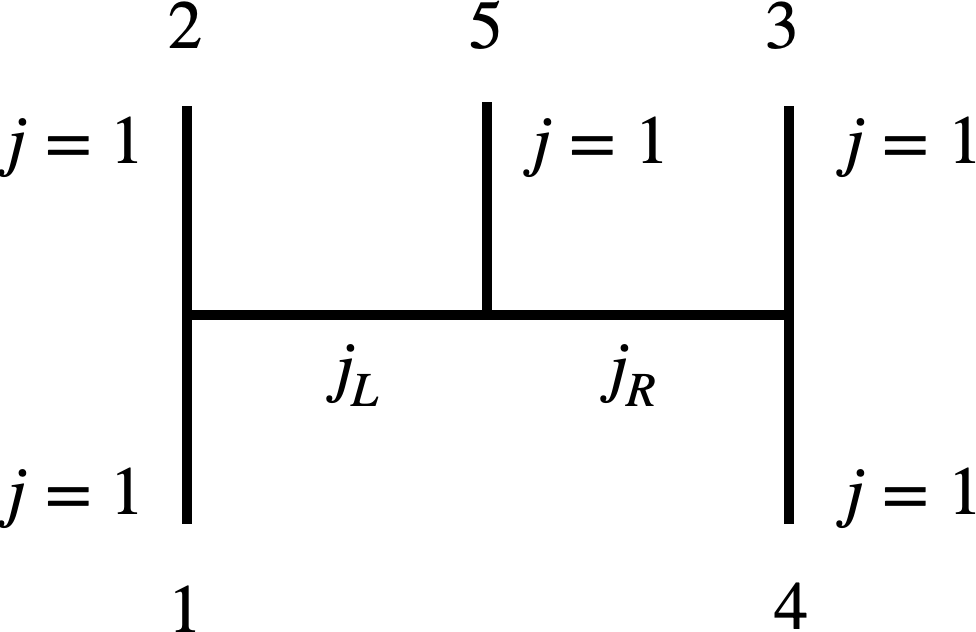}
\caption{Fusion of five spin-1 $SU(2)$ representations. The possible intermediate spins $(j_L,j_R)$ are (0,1), (1,0), (1,1), (1,2), (2,1), (2,2).}
    \label{fig:SU2R}
\end{figure}

\subsubsection{$SU(2)_R$ structures}
Covariance under $SU(2)_R$ requires polarization spinors to appear  in the correlation function as polynomials of $\prod_{i<j}(v_i\cdot v_j)^{a_{ij}}$ with $a_{ij}=a_{ji}\geq 0$, $a_{ii}=0$ and $\sum_{j}a_{ij}=2$. These monomials  satisfy linear relations and there are only six independent elements. A convenient basis can be chosen such that the polynomials correspond to six possible fusion processes in fig. \ref{fig:SU2R}
\begin{eqnarray}
\nonumber T_{0,1}=&&-V_{13245}+V_{13254}-V_{13425}+V_{13524}\\
\nonumber &&+V_{14235}+V_{14325}\;,\\
\nonumber T_{1,0}=&&V_{13425}+V_{14325}\;,\\
\nonumber T_{1,1}=&&V_{13245}+V_{13254}+V_{13524}-V_{14235}\;,\\
\nonumber T_{1,2}=&&V_{13245}+V_{13254}+\tfrac{2}{3}V_{13425}-V_{13524}\\
\nonumber &&+V_{14235}+\tfrac{2}{3}V_{14325}\;,\\
\nonumber T_{2,1}=&&-V_{13245}+V_{13254}+2V_{13425}+V_{13524}\\
\nonumber &&+V_{14235}-2V_{14325}\;,\\
\nonumber T_{2,2}=&&V_{13245}-V_{13254}+V_{13524}+V_{14235}\;.
\end{eqnarray}
These polynomials are eigenfunctions of the bi-particle Casimirs of $SU(2)_R$
\begin{equation}
\begin{split}
\nonumber {\rm Cas}_{12}T_{j_L,j_R}=j_L(j_L+1)T_{j_L,j_R}\;,\\
{\rm Cas}_{34}T_{j_L,j_R}=j_R(j_R+1)T_{j_L,j_R}\;.
\end{split}
\end{equation}
In terms of these R-symmetry structures, we can parameterize $Y_X(g,v)$ in the ansatz as 
\begin{eqnarray}
\nonumber Y_{s,s}(g,v)=&&\lambda_{s,s} g_7 T_{1,1}\;,\\
Y_s(g,v)=&&\sum_{a=0}^2\lambda_s^{(1,a)}g_7 T_{1,a}+\sum_{a=0}^2\lambda_s^{(2,a)}g_6T_{1,a}\\
\nonumber &&+(\text{permutations in 345})\;.\\
\nonumber Y_{v,s}(g,v)=&&\lambda_{v,s}g_7T_{1,0}\;.
\end{eqnarray}
There appears to be six parameters in $Y_s$. However, using the relations among the structures we find that there is only one after symmetrization with respect to $345$.

\bibliography{refg5pt} 
\bibliographystyle{utphys}
\end{document}